\documentclass[prb,twocolumn]{revtex4-1}

\usepackage{amsmath,amsfonts,amssymb}
\usepackage{graphicx,color}
\usepackage{hyperref}
\usepackage{epstopdf}
\usepackage{dcolumn}   
\usepackage{array,multirow,graphicx}
\begin{document}
	
	\title{Engineering Cross Resonance Interaction in Multi-modal Quantum Circuits}
	\author{Sumeru Hazra$^{1}$}
	\author{Kishor V. Salunkhe$^{1}$, Anirban Bhattacharjee$^{1}$, Gaurav Bothara$^{1}$, Suman Kundu $^{1}$}
	\author{Tanay Roy$^{1}$}
	\altaffiliation{Current address: Department of Physics and James Franck Institute, University of Chicago, Chicago, Illinois 60637, USA}
	\author{Meghan P. Patankar$^{1}$}
	\author{R. Vijay$^{1}$}
	\email{Corresponding author: r.vijay@tifr.res.in}
	\affiliation{$^{1}$
		Department of Condensed Matter Physics and Materials Science,\\
		Tata Institute of Fundamental Research,
		Homi Bhabha Road, Mumbai 400005, India
	}
	\date{\today}
	
	\begin{abstract}
		Existing scalable superconducting quantum processors have only nearest-neighbor coupling. This leads to reduced circuit depth, requiring large series of gates to perform an arbitrary unitary operation in such systems. Recently, multi-modal devices have been demonstrated as a promising candidate for small quantum processor units. Always on longitudinal coupling in such circuits leads to implementation of native high fidelity multi-qubit gates. We propose an architecture using such devices as building blocks for a highly connected larger quantum circuit. To demonstrate a quantum operation between such blocks, a standard transmon is coupled to the multi-modal circuit using a 3D bus cavity giving rise to small exchange interaction between the transmon and one of the modes. We study the cross resonance interaction in such systems and characterize the entangling operation as well as the unitary imperfections and cross-talk as a function of device parameters. Finally, we tune up the cross resonance drive to implement multi-qubit gates in this architecture.
	\end{abstract}
	
	\maketitle
	Superconducting qubits have become one of the most promising platforms for quantum computation and quantum information processing\cite{chuang_book} in the near term. Over the past decade small quantum processors with superconducting qubits have shown tremendous improvement in terms of coherence times reaching milliseconds\cite{high_coherence_fluxonium, millisec_mem} and scalability up to 10-70 qubits\cite{quantum_supremacy, ibmq}. However almost all the existing architectures\cite{google_chip, ibm_surface, ibm_ft, eth_multiplex} in superconducting qubits have only nearest-neighbor coupling. With limited connectivity this often imposes strong constraints on available multi-qubit operations in such architectures and leads to inefficient implementation of quantum algorithms and quantum simulations\cite{connectivity}. On the other hand, always on all-to-all interaction in longitudinally coupled multi-modal devices\cite{trimon1, trimon_th} leads to implementation of fast high fidelity N-qubit gates in the circuit. Previous experiments have demonstrated such devices as an effective three qubit processor with efficient implementation of small quantum algorithms\cite{trimon2}. Using multi-modal devices as building blocks for a larger quantum processor could enable greater interqubit connectivity and increased circuit depth for quantum information processing. This is a useful approach to enhance the performance of near-term imperfect quantum processors\cite{nisq, vaq} without fault tolerance.
	
	In this letter, we demonstrate a circuit QED architecture consisting of a multi-modal superconducting circuit\cite{trimon_th} and a transmon \cite{transmon} qubit coupled via an exchange $(\sigma_x\sigma_x)$ coupling mediated via a bus cavity. We numerically analyze the effect of a cross resonance\cite{cr_rigetti, cr_chow} drive in such systems and estimate the elements of the effective Hamiltonian \cite{intrinsic_error_budget} for experimentally realizable parameters. Then we use a frequency tunable transmon to experimentally study the cross resonance effect as a function of detuning between the two qubits. We identify the optimum detuning range and tune up cross resonance interaction for a multi-qubit entangling gate. We characterize the performance of the gate using randomized benchmarking protocol as well as via quantum process tomography. We propose a scalable quantum computing architecture using multi-modal circuits as building blocks with enhanced circuit depth.
	
	\begin{figure}[h]
		\centering
		\includegraphics[width=0.45\textwidth]{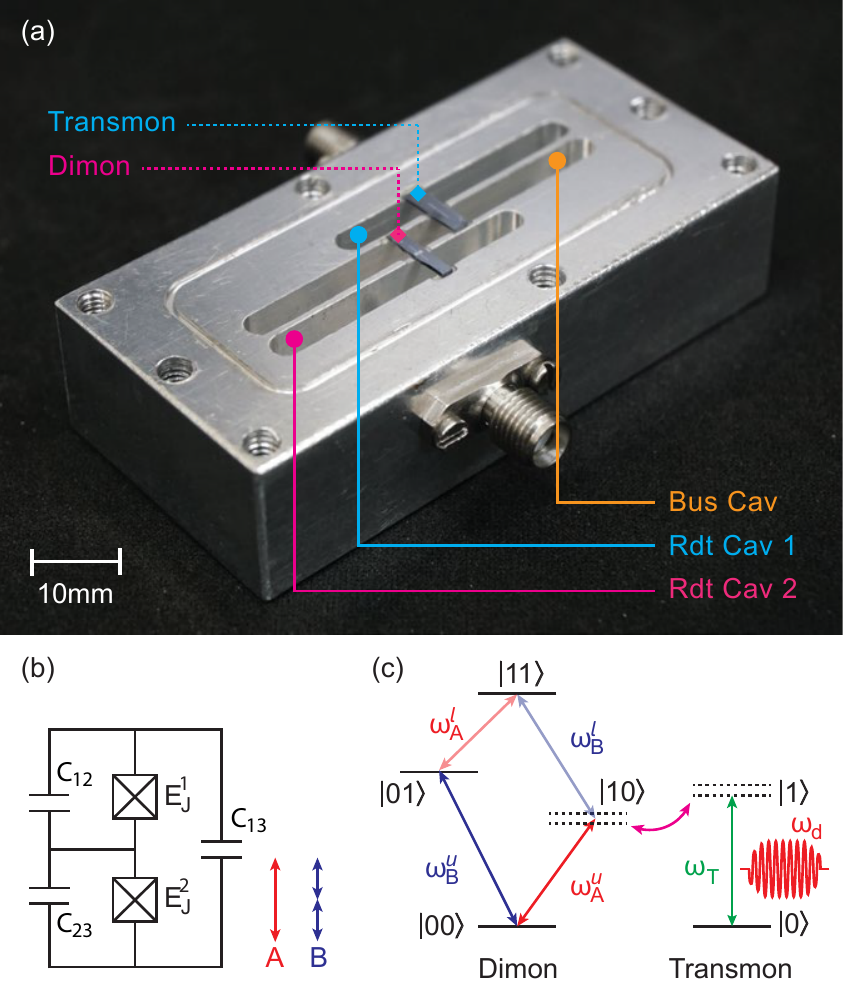} 
		\vspace{10pt}
		\caption{(a) Multi-modal circuit dimon and transmon in the 3D aluminum bus cavity architecture. The qubits are shared between the bus resonator and the dedicated readout resonator for qubit measurement. The bus cavity mediates the exchange coupling for cross resonance interaction. (b) Lumped element circuit equivalent for dimon. A and B are the two normal modes of the circuit, respectively the dipolar mode and the quadrupolar mode. (c) Energy level diagram of the computational subspace of the three-qubit Hilbert space. Due to the longitudinal coupling each mode k of dimon has two transition frequencies $\omega^u_k$ and $\omega^l_k$ depending on the state of its partner qubit. The $|0\rangle$ to $|1\rangle$ transition frequency of the transmon is denoted as $\omega_{T}$. For cross resonance interaction the transmon is used as the control qubit and it is driven at the upper sideband frequency $(\omega^u_A)$ of the A mode. The dashed lines indicate the weak hybridization due to the transverse coupling of the transmon with the A mode of dimon.}
		\label{fig:circuit}
	\end{figure}
	
	In this work, we have chosen a two-qubit version of the multi-modal circuit \cite{trimon_th} to demonstrate the CR gate. This two-mode circuit consists of two Josephson junctions in series \cite{houck_dimon} as shown in Fig. \ref{fig:circuit}.(b) and is named as dimon. Each qubit has two transition frequencies depending on the state of its partner qubit allowing one to implement a fast high fidelity controlled operation \cite{trimon1} in the two-qubit subspace (Fig.~\ref{fig:circuit} (c)) . For single qubit rotations one needs to drive both transitions of the intended qubit. For readout, one of the modes is coupled to the electric field of a resonator. Due to the longitudinal coupling, both the modes dispersively shift the resonator enabling standard joint readout of the two-qubit system\cite{trimon1}.
	
	In a typical cross resonance (CR) architecture \cite{cr_rigetti, cr_chow, cr_leek} two frequency detuned transmons are weakly hybridized via a fixed transverse coupling. One of the qubits (control) is driven at the transition frequency of the other qubit (target). As a result the latter starts coherently oscillating between the $|0\rangle$ and $|1\rangle$ levels. The rate of such evolution depends on the state of control qubit, thus enabling an entangling operation in the two-qubit subspace.
	
	In our implementation, we replace one of the transmons by a dimon. We write down the Hamiltonian for the system as:
	\begin{equation}
	\label{hm1}
	\hat{\mathcal{H}}_{\rm{sys}}=\hat{\mathcal{H}}_0+\hat{\mathcal{H}}_D
	\end{equation}
	where, $\hat{\mathcal{H}}_0$ is the bare Hamiltonian with longitudinal coupling between the qubits (A and B) of the multi-modal system and transverse coupling between transmon (T) and the dipolar mode (A) of dimon,
	\begin{equation}
	\label{hm2}
	\begin{split}
	\hat{\mathcal{H}}_0/\hbar=&\sum_{i=A,B,T}((\omega_i-\beta_i)\hat{a}^\dagger_i\hat{a}_i-\delta_i(\hat{a}^\dagger_i\hat{a}_i)^2)\\&+2J_L\hat{a}^\dagger_A\hat{a}_A\hat{a}^\dagger_B\hat{a}_B+J_T(\hat{a}^\dagger_A\hat{a}_T+\hat{a}^\dagger_T\hat{a}_A)
	\end{split}
	\end{equation}
	and $\hat{\mathcal{H}}_D$ represents the cross resonance drive term applied on the transmon (control qubit).
	\begin{equation}
	\label{cr_drive}
	\begin{split}
	\hat{\mathcal{H}}_D=&\Omega_d(t)\cos(\omega_dt+\phi_T)(\hat{a}^\dagger_T+\hat{a}_T)\\&+m\Omega_d(t)\cos(\omega_dt+\phi_A)(\hat{a}^\dagger_A+\hat{a}_A)
	\end{split}
	\end{equation}
	where $\Omega_d(t)$ is the amplitude of the cross resonance drive and $\omega_d$ is the drive frequency. The second term represents the classical cross-talk which couples to the target qubit and can induce direct Rabi oscillations. $\phi_T$ and $\phi_A$ are the respective phases of the drive acting on the two qubits. $\delta_i$ is anharmonicity of the i\textsuperscript{th} qubit and $\hat{a}_i$ is the annihilation operator acting on i\textsuperscript{th} oscillator states. In the first experiment we have used a fixed amplitude square pulse for CR drive and $\omega_d$ is chosen to be equal to the upper sideband transition frequency $\omega_A^u$ of the A mode of dimon. Since this is detuned from the lower sideband transition frequency $\omega_A^l$ of the dimon, the CR drive results in controlled rotation in the $|00\rangle$ and $|10\rangle$ subspace of dimon depending on the state of the transmon and leaves the $|01\rangle$ and $|11\rangle$ states unaffected. Consequently in the full three-qubit subspace this implements a three-qubit gate of Toffoli class. Note that, in our implementation of cross resonance we have chosen the transmon as the control qubit because it has a higher anharmonicity. The entangling gate in the reverse direction can be constructed with additional single qubit rotations on both qubits.

	\begin{table*}[t]
		\begin{tabular}{c c l c c c c c c c c} 
			\hline
			\hline
			\\[-2ex]
			& & & Cavity & Cavity & Qubit & Anharmoni-   &  &  Coherence &    \\
			& & Qubit & frequency & line width & frequency* & city  & T$_1$ & $\rm{T}^2_{\rm{Ramsey}}$ & $\rm{T}^2_{\rm{Echo}}$  \\
			& & & ${\omega_R}/{2\pi}$ (GHz) & ${\kappa_R}/{2\pi}$ (MHz) & ${\omega_q}/{2\pi}$ (GHz) & ${\delta}/{2\pi}$ (MHz)  & ($\mu$s) & ($\mu$s) & ($\mu$s) \\
			\hline
			\hline
			\\[-2ex]
			\parbox[t]{2mm}{\multirow{3}{*}{\rotatebox[origin=c]{90}{EXP1}}} &\parbox[t]{2mm}{\multirow{3}{*}{\rotatebox[origin=c]{90}{(Cu)}}} & Dimon A mode & 7.340 & 3.315 & 4.413 & -100  & 10 & 1.6  & 1.8 &\\ 
			& & Dimon B mode & 7.340 & 3.315 & 5.620 & -123  & 5.2 & 1.3 & 2.1 &\\
			& & Split junction Transmon & 7.287 & 4.238 & 4.959 & -220   & 11.3 & 1.1 & 1.6 &\\ [0.5ex]
			\hline
			\\[-2ex]
			\parbox[t]{2mm}{\multirow{3}{*}{\rotatebox[origin=c]{90}{EXP2}}} & \parbox[t]{2mm}{\multirow{3}{*}{\rotatebox[origin=c]{90}{(Al)}}}& Dimon A mode & 7.307 & 3.785 & 4.562 & -128  & 18 & 17  &19&\\ 
			& & Dimon B mode & 7.307 & 3.785 & 5.822 & -142  & 7 & 4 & 4&\\
			& & Fixed frequency Transmon & 7.276 & 3.158 & 4.774 & -280   & 14 & 6 & 8 &\\ [0ex]
			\hline
			\hline
		\end{tabular}
		\caption{Measured device parameters and coherence numbers of the multi-modal circuit and the transmon in their corresponding readout cavities used in the two experiments in copper and aluminum cavity respectively. In the first experiment tabulated transmon parameters correspond to zero flux condition. (*upper sideband frequency in case of dimon)}
		\label{table:deviceparameter}
	\end{table*}
	
	First, we investigate the dependence of the cross resonance interaction on interqubit detuning\cite{tunable_cr}. The experimental setup consists of three rectangular waveguide copper cavities, a geometry identical to Fig. \ref{fig:circuit}. The bus cavity\cite{bus_cav} at the center has a frequency 6.4GHz. The two readout cavities are used to measure the qubit states as well as to send microwave tones to manipulate the qubit states. The transmon and the multi-modal circuit are shared between the bus cavity and individual readout cavities. The capacitor pads are extended like an antenna to couple each qubit with the TE101 mode of their respective cavities. We use a split junction transmon as a frequency tunable qubit, controllable by an external flux bias through the copper cavity. The resonant frequencies of the readout resonators were measured to be $\omega_{R1}/2\pi= 7.287$ $\rm{GHz}$ and $\omega_{R2}/2\pi= 7.340$ $\rm{GHz}$ when the transmon is biased at its zero flux condition. We perform two-tone spectroscopy to extract the relevant transition frequencies of the system. Device parameters of the multi-qubit cavity system are mentioned in the first part of table \ref{table:deviceparameter}. The longitudinal coupling between the two dimon modes is $J_L/2\pi = 70.5$ $\rm{MHz}$. The bare coupling between the dipolar mode of dimon and the transmon is measured to be equal to $J_T/2\pi = 1.9$ $\rm{MHz}$ from avoided crossing when the transmon is tuned onto resonance with the A mode upper sideband transition.
	
	Next we park the tunable transmon at some specific detuning values with respect to $\omega_A^u$. Due to the presence of higher levels of both qubits the exchange coupling gives rise to small effective static ZZ interaction in the two-qubit subspace comprised of the A mode of the dimon and the transmon. This interaction further splits the dimon upper sideband transition (target qubit) frequency into $\omega_{A0}^u$ and $\omega_{A1}^u$ . We measure this splitting by performing a pair of conditional Ramsey experiments where the transmon (control qubit) is kept at $|0\rangle$ and $|1\rangle$ states respectively. The frequency of the cross resonance drive in Eq. \ref{cr_drive} is chosen to be the mean of these two frequencies, $\omega_d=(\omega_{A0}^u+\omega_{A1}^u)/2$.
	
	We perform a Hamiltonian tomography \cite{hamiltonian_tomography_ibm} experiment where we apply the cross resonance drive on the transmon and the length of this pulse is varied from 0 to $\tau_{max}$ like a Rabi experiment. This is followed by tomographic measurements on the target qubit. We carry out two experiments with the control qubit in the ground and excited state respectively. From the two sets of qubit state evolution, we extract the elements of the effective block diagonal Hamiltonian consisting of ZX, ZY, ZZ, IX, IY and IZ. We adjust the phase of the cross resonance drive with respect to the target qubit axes such that the $\rm{ZX}$ component is maximized while minimizing the $\rm{ZY}$ interaction strength\cite{supp}.
	
	\begin{figure}[b]
		\centering
		\includegraphics[width=0.5\textwidth]{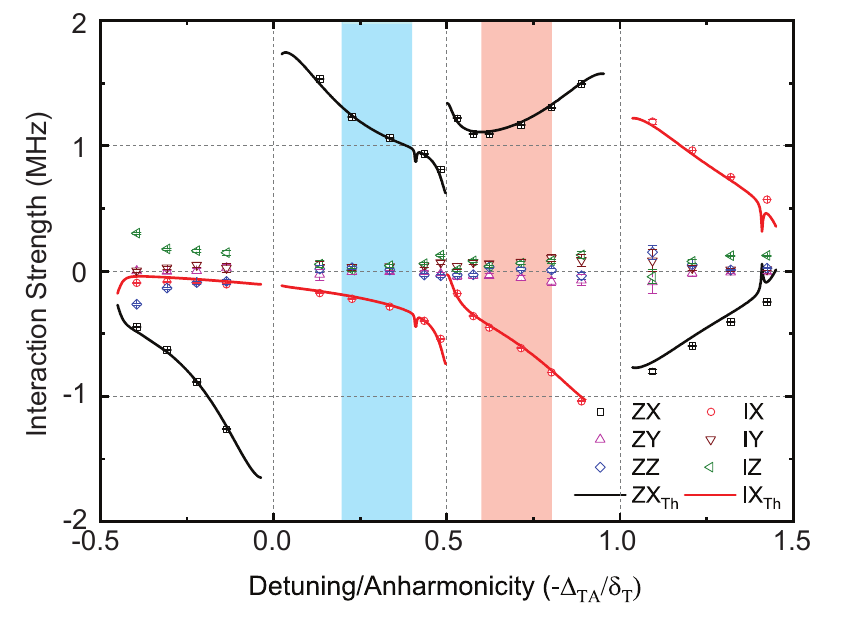} 
		\vspace{-5pt}
		\caption{Interaction strengths as a function of dimensionless detuning $-\Delta_{\rm{TA}}/\delta_T$ (normalized by the anharmonicity of the control qubit). The interaction strengths are extracted from Hamiltonian tomography on the multi-qubit architecture with transmon as the control qubit and the upper sideband transition of the dipolar mode of dimon as the target qubit. The transmon is parked at different frequencies by varying the flux bias to sweep the interqubit detuning from  $-\delta_T/2$ to $+3\delta_T/2$. The theoretical plot is extracted from an effective Hamiltonian theory with higher levels. The two optimal ranges of detuning for cross resonance are shaded in light blue and pink respectively.}
		\label{fig:hamiltonian_tomography}
	\end{figure}
	
	We vary the interqubit detuning $\Delta_{TA}=\omega_T-\omega_A^u$ from a negative value of $-\delta_T/2$ to a positive value of $+3\delta_T/2$ where $\delta_T$ is the anharmonicity of the transmon. We perform Hamiltonian tomography at each detuning to find the dependence  of the elements of the effective Hamiltonian on interqubit detuning (Fig.\ref{fig:hamiltonian_tomography}). The dependence is accurately modeled using anharmonic oscillators containing higher levels \cite{supp}. A comparison of the experimental result with the quantitative plot generated by numerical estimation is shown in Fig.\ref{fig:hamiltonian_tomography}.  When the detuning between the two qubits approaches $-\delta_T/2$, $0$, $\delta_T/2$, $\delta_T$ and $3\delta_T/2$, the cross resonance pulse drives unwanted transitions on the control qubit\cite{ibm_eff_ham}. These values of detuning are avoided while optimizing two-qubit gates as they would generate strong leakage out of the computational subspace. From the plot we observe that for a given exchange coupling between the qubits and for a constant drive strength, the ZX interaction is larger for positive detuning. Also ZX interaction falls rapidly for detuning larger than anharmonicity because the weakly anharmonic oscillators appear to be more like linear systems to each other when far detuned \cite{intrinsic_error_budget,ibm_eff_ham}. We also notice that the ZZ interaction strength, that is the leading non-commuting error term, gets minimized in the range $0<\Delta_{\rm{TA}}<\delta_T$. The constant offset in the ZZ term is due to static ZZ split. There is an IX term arising from the presence of the higher levels. ZY and IY terms are simultaneously minimized by adjusting the phase of the cross resonance drive, indicating negligible classical cross-talk in our implementation. This is because in 3D geometry the readout resonators are well isolated from each other and from the bus cavity, offering better qubit addressability. This is in contrast to experimental results obtained in 2D architectures\cite{hamiltonian_tomography_ibm} where a large IY term was reported and an active cancellation pulse was required. Based on the study of interaction terms as a function of detuning we observe that the best operational regimes for cross resonance gate in terms of minimum leakage and maximum speed are $0.2<-\Delta_{TA}/\delta_T<0.4$ and $0.6<-\Delta_{TA}/\delta_T<0.8$, as it has been theoretically predicted \cite{intrinsic_error_budget} for the case of two transmons.
	
	In the next experiment we use a fixed frequency transmon and a dimon in an aluminum cavity with optimum parameters for cross resonance. The set up is shown in Fig. \ref{fig:circuit}.(a).	The corresponding qubit parameters and coherence numbers are displayed in the second half of Table \ref{table:deviceparameter}. We carry out a conditional Ramsey experiment and measure a static ZZ splitting ($2\xi/2\pi$) of 168 kHz between the transversely coupled qubits. We estimate a coupling strength ($J_T/2\pi$) of 2.76 MHz using the relation\cite{ibm_eff_ham}:
	\begin{equation}
	\label{ZZ splitting}
	2\xi= -\frac{2J_T^2(\delta_T+\delta^u_A)}{(\Delta_{\rm{TA}}+\delta_T)(\delta^u_A-\Delta_{\rm{TA}})}
	\end{equation}

	\begin{figure}[t]
		\centering
		\includegraphics[width=0.5\textwidth]{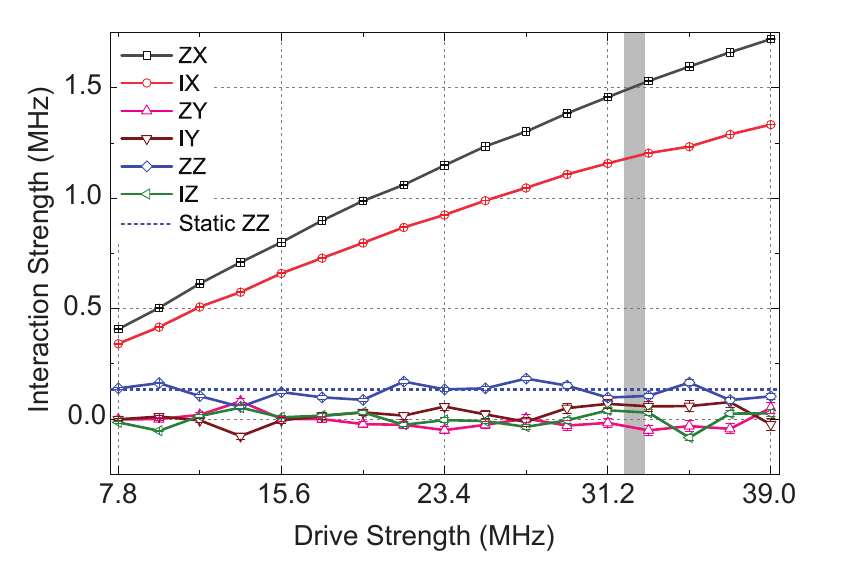} 
		\vspace{-5pt}
		\caption{Amplitude of interaction terms in the effective Hamiltonian as a function of cross resonance drive strength measured in the second experiment with fixed frequency qubits in aluminum cavity. Phase of the cross resonance drive is optimized to minimize the ZY term. The drive amplitude used in the CR gate is shaded in gray.}
		\label{fig:ht_vs_drive}
	\end{figure}
	
\begin{figure}[t]
		\centering
		\includegraphics[width=0.5\textwidth]{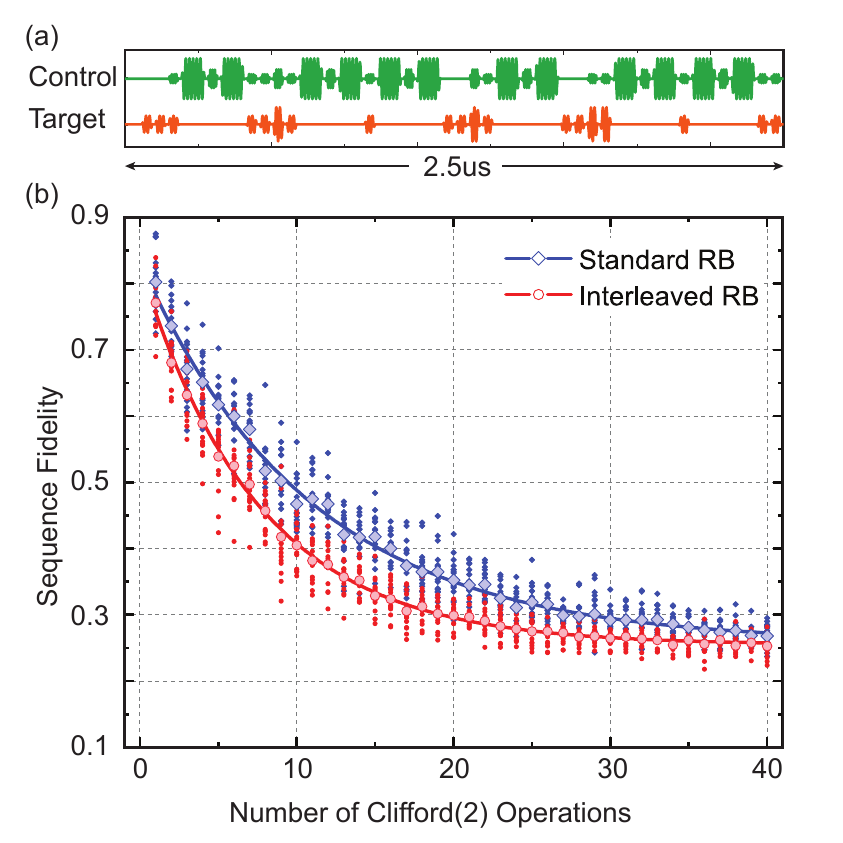} 
		\vspace{-0pt}
		\caption{(a) A random two-qubit pulse sequence for the RB experiment, consisting of 3 randomly chosen operations from Clifford(2) group. Tall Gaussian pulses represent $\pi$ rotations, short ones correspond to $\pi/2$ rotations. Both $\pi$ and $\pi/2$ operations are 40 ns in length. Longer pulses on the control qubit represent cross resonance drive. (b) Standard and interleaved randomized benchmarking to characterize the cross resonance gate. We estimate an average two qubit gate fidelity of $0.9344\pm0.0019$ from the standard protocol and a gate fidelity of $0.9703\pm0.0037$ for the CR operation from the interleaved RB protocol.}
		\label{fig:rb_and_qpt}
	\end{figure}

	The single qubit gates in the system are comprised of 40 ns rounded square pulses which include 10 ns Gaussian rise and fall time. Successive pulses are spaced by 5 ns delays. Average single qubit gate fidelities are $0.9973\pm 0.0001$ and $0.9959\pm 0.0002$ respectively for the dimon and the transmon, measured with simultaneous randomized benchmarking. Next we tune up the two-qubit gate by first measuring the effective interaction strengths as a function of cross resonance drive amplitude. The dependence is shown in the Fig. \ref{fig:ht_vs_drive}. We calibrate the drive amplitude using the direct Rabi drive strength applied on the control qubit. The drive strength used in the calibrated gate is shaded in gray. The phase of cross resonance drive is chosen to minimize ZY interaction. Due to negligible classical cross talk in this architecture, the same added phase also cancels the IY term in the effective Hamiltonian. Next we perform echoed cross resonance \cite{CR_RB} that eliminates the IX term and the stark shift term (ZI) that arise from driving the control qubit. The echoed $\rm{CR}$ sequence consists of two cross resonance pulses with opposite signs separated by a $\pi$ pulse on control qubit. This implements a $\rm{ZX}_{\pi/2}$ operation in the two-qubit subspace up to a bit-flip on the control qubit. The total gate time for the echoed $\rm{CR}$  operation is 220 ns consisting of two cross resonance pulses of 85 ns each and a $\pi$ pulse of 40 ns spaced by a gap of 5 ns on either side. The $\rm{CR}$ pulses and the echo pulse are rounded square pulses with Gaussian rise and fall time of 10 ns \cite{supp}. We characterize the performance of the gate by interleaved randomized benchmarking (RB) \cite{interleaved_rb}. We first perform a standard RB experiment with randomly generated Clifford sequences each with a maximum length of 40 two-qubit operations. We fit the sequence fidelity to a decaying exponential averaged over 17 random sequences as shown in Fig. \ref{fig:rb_and_qpt}(b). Next we interleave the target gate with the same random Clifford sequence and find another decay curve. From the two decay constants we estimate the fidelity of the $\rm{CR}$ gate to be $0.9703\pm 0.0037$. We have also performed quantum process tomography\cite{chuang_book} of the  $\rm{ZX}_{\pi/2}$ gate as an independent benchmark. We report a process fidelity of $0.9282\pm0.0002$ and a gate fidelity of $0.9426\pm0.0002$ from quantum process tomography \cite{supp}. While performing QPT the states were corrected for measurement error. The measurement fidelities for the transmon and the dimon were 0.930 and 0.938 respectively and we used heralding to initialize the system in the ground state. Primary contribution to the gate infidelity is due to decoherence of the qubits. The coherence limit for CR gate is estimated\cite{supp} to be 0.9787. The small coherent error term in the cross resonance gate is mainly attributed to the ZZ error and leakage out of subspace. 
	
	In conclusion, we have adapted the standard cross resonance interaction to implement a multi-qubit operation between a transmon and a longitudinally coupled two-qubit circuit in a 3D cQED architecture. Any of the sideband transition frequencies of the multi-modal circuit may be used as the ``target" in the CR scheme, implementing a CCNOT or Toffoli gate. The scheme outlined here could be generalized to N-qubit multi-modal circuits implementing a N-qubit Toffoli gate. Finally, this architecture can be scaled to couple multiple blocks of all-to-all coupled multi-modal circuits to build a small scale quantum processor with enhanced interqubit connectivity.
	
	\textit{Acknowledgments:-} This work was supported by the Department of Atomic Energy of the Government of India. R.V. acknowledges support from the Department of Science and Technology, India via the Nano Mission. We acknowledge the TIFR nanofabrication facility.

%

	\newpage
	\setcounter{figure}{0}
	\setcounter{table}{0}
	\setcounter{equation}{0}
	\newpage
	\global\long\def\theequation{S\arabic{equation}}
	\global\long\def\thefigure{S\arabic{figure}}
	\global\long\def\thetable{S\arabic{table}}
	\onecolumngrid
	\begin{center}
		{\bf \large Supplementary Information for ``Engineering Cross Resonance Interaction in Multi-modal Quantum Circuits"}
	\end{center}

	\begin{center}	
		{Sumeru Hazra$^{1}$, Kishor V. Salunkhe$^{1}$, Anirban Bhattacharjee$^{1}$, Gaurav Bothara$^{1}$, Suman Kundu $^{1}$, Tanay Roy$^{1,2}$\\
		Meghan P. Patankar$^{1}$ and R. Vijay$^{1}$}\\
		$^{1}$\textit{Department of Condensed Matter Physics and Materials Science, \\ Tata Institute of Fundamental Research, Homi Bhabha Road, Mumbai 400005, India and\\ $^{2}$Department of Physics and James Franck Institute, \\ University of Chicago, Chicago, Illinois 60637, USA}
	\end{center}
	\twocolumngrid
	\section{Numerical Analysis of Cross Resonance in Multimodal Systems}
	In order to numerically find the effective block diagonal Hamiltonian we closely follow the approach outlined in \cite{ibm_th}. We begin with the duffing oscillator Hamiltonian defined in the main text
	\begin{equation}
	\label{hm1}
	\hat{\mathcal{H}}_{\rm{sys}}=\hat{\mathcal{H}}_0+\hat{\mathcal{H}}_D
	\end{equation}
	where the bare Hamiltonian $\hat{\mathcal{H}}_0$ contains the three non-linear oscillator terms corresponding to the transmon $(T)$ and the two modes $(A,B)$ of dimon and the transverse $(J_{\rm{xx}})$ and longitudinal $(J_{\rm{zz}})$ coupling between respective qubits.
	\begin{equation}
	\label{hm2}
	\begin{split}
	\hat{\mathcal{H}}_0/\hbar=&\sum_{i=T,A,B}\left((\omega_i-\beta_i)\hat{a}^\dagger_i\hat{a}_i-\delta_i(\hat{a}^\dagger_i\hat{a}_i)^2\right)\\&+2J_{\rm{zz}}\hat{a}^\dagger_A\hat{a}_A\hat{a}^\dagger_B\hat{a}_B+J_{\rm{xx}}\left(\hat{a}^\dagger_A\hat{a}_T+\hat{a}^\dagger_T\hat{a}_A\right)\\&+\lambda J_{\rm{xx}}\left(\hat{a}^\dagger_B\hat{a}_T+\hat{a}^\dagger_T\hat{a}_B\right)
	\end{split}
	\end{equation}
	With perfectly identical junctions one mode of the dimon is dipolar and the other mode is quadrupolar. Hence only one of the modes is coupled to the transmon. However due to asymmetry between the junctions effective modes have both dipolar and quadrupolar components depending on the degree of mode mixing. Typically the junction asymmetry is kept small and $\lambda(<<1)$ is the resulting relative coupling of the $B$ mode compared to the $A$ mode.
	
	$\hat{\mathcal{H}}_D$ contains the cross resonance drive applied on the control transmon and a smaller drive term on the target qubit due to classical cross-talk. The phase of the cross-talk term in general could be different from that of the cross resonance drive.
	\begin{equation}
	\label{cr_drive}
	\begin{split}
	\hat{\mathcal{H}}_D=&\Omega_d\cos(\omega_dt+\phi_d)(\hat{a}^\dagger_T+\hat{a}_T)\\&+m\Omega_d\cos(\omega_dt+\phi'_d)(\hat{a}^\dagger_A+\hat{a}_A)\\&+m\lambda\Omega_d\cos(\omega_dt+\phi'_d)(\hat{a}^\dagger_B+\hat{a}_B)
	\end{split}
	\end{equation}
	where $\Omega_d$, $\omega_d$ and $\phi_d$ are the amplitude, frequency and phase of the cross resonance drive respectively, while $m$ is some small fraction quantifying the degree of cross-talk. Since $m\lambda<<1$ we drop the last term in equation \ref{cr_drive}. We define the ladder operators for the three duffing oscillators upto $d$ levels and arrange them as  $\left\lbrace 000, 001, 010, 100, 002, 011, 020, 101, 110, 200, ...,d00\right\rbrace$.
	
	We first rotate the bare Hamiltonian $\hat{\mathcal{H}}_0$ into its diagonal form $\hat{\mathcal{H}}^{\rm{diag}}_0$, by applying the unitary $\hat{\mathcal{U}}$ composed of the eigenvector matrix of $\hat{\mathcal{H}}_0$.
	\begin{equation}
	\label{diag}
	\hat{\mathcal{H}}^{\rm{diag}}_0=\hat{\mathcal{U}}^\dagger\hat{\mathcal{H}}_0 \hat{\mathcal{U}}
	\end{equation}
	The drive term of Eq.\ref{cr_drive} is then rotated into this frame using the same unitary $\hat{\mathcal{U}}$.
	\begin{equation}
	\label{diag_drive}
	\begin{split}
	\hat{\mathcal{H}}_D^{\rm{diag}}=&\Omega_d\cos(\omega_dt+\phi_d)\hat{\mathcal{U}}^\dagger(\hat{a}^\dagger_T+\hat{a}_T)\hat{\mathcal{U}}\\&+m\Omega_d\cos(\omega_dt+\phi'_d)\hat{\mathcal{U}}^\dagger(\hat{a}^\dagger_A+\hat{a}_A)\hat{\mathcal{U}}
	\end{split}
	\end{equation}
	Breaking the cosine into exponential form and writing $\hat{\mathcal{D}}_j=\hat{\mathcal{U}}^\dagger(\hat{a}^\dagger_j+\hat{a}_j)\hat{\mathcal{U}}$ we get,
	\begin{equation}
	\label{diag_drive2}
	\begin{split}
	\hat{\mathcal{H}}_D^{\rm{diag}}=\Omega_d\cos{\phi_d}&\left( \frac{e^{i\omega_dt}+e^{-i\omega_dt}}{2}\right)\hat{\mathcal{D}_T}\\+i\Omega_d&\sin{\phi_d}\left( \frac{e^{i\omega_dt}-e^{-i\omega_dt}}{2}\right)\hat{\mathcal{D}_T}\\+&m\Omega_d\cos{\phi'_d}\left( \frac{e^{i\omega_dt}+e^{-i\omega_dt}}{2}\right)\hat{\mathcal{D}_A} \\&+im\Omega_d\sin{\phi'_d}\left( \frac{e^{i\omega_dt}-e^{-i\omega_dt}}{2}\right)\hat{\mathcal{D}_A}
	\end{split}
	\end{equation}
	At this point we rotate all the qubits in a frame rotating at the drive frequency $\omega_d$ and apply the RWA. We denote this frame transformation by a unitary operator $\hat{\mathcal{R}}$ acting on $	\hat{\mathcal{H}}_{\rm{sys}}^{\rm{diag}}$ defined as:
	\begin{equation}
	\label{RWA}
	\hat{\mathcal{R}}=e^{-i\hat{\mathcal{H}}_{\rm_{rot}}t}=e^{-i\omega_d\left(\hat{a}^\dagger_A\hat{a}_A+\hat{a}^\dagger_B\hat{a}_B+\hat{a}^\dagger_T\hat{a}_T\right)t}
	\end{equation}
	
	The Hamiltonian in this rotating frame is given by,
	\begin{equation}
	\label{RWA}
	\begin{split}
	\hat{\mathcal{H}}_{\rm{sys}}^{\rm{RWA}}&=\hat{\mathcal{R}}^\dagger\hat{\mathcal{H}}_{\rm{sys}}^{\rm{diag}} \hat{\mathcal{R}}-i\hat{\mathcal{R}}^\dagger\dot{\hat{\mathcal{R}}}\\
	&=\hat{\mathcal{R}}^\dagger\hat{\mathcal{H}}_{0}^{\rm{diag}}\hat{\mathcal{R}}+\hat{\mathcal{R}}^\dagger\hat{\mathcal{H}}_{\rm{D}}^{\rm{diag}}\hat{\mathcal{R}}-i\hat{\mathcal{R}}^\dagger\dot{\hat{\mathcal{R}}}\\
	&=\hat{\mathcal{H}}_{0}^{\rm{diag}}-\hat{\mathcal{H}}_{\rm_{rot}}+\hat{\mathcal{R}}^\dagger\hat{\mathcal{H}}_{\rm{D}}^{\rm{diag}}\hat{\mathcal{R}}
	\end{split}
	\end{equation}
	We first evaluate the term, $\hat{\mathcal{R}}^\dagger\hat{\mathcal{H}}_{\rm{D}}^{\rm{diag}}\hat{\mathcal{R}}$ under RWA and ignore terms with energy cost $\omega_d$ or higher. The Hamiltonian $	\hat{\mathcal{H}}_{\rm{sys}}^{\rm{RWA}}$ captures the full dynamics of the system including high frequency oscillations on top of the slow envelope of CR evolution in the target qubit due to leakage in the control qubit. 
	
	Next we find an effective Hamiltonian using the principle of least action\cite{least_action} from the RWA Hamiltonian $\hat{\mathcal{H}}_{\rm{sys}}^{\rm{RWA}}$. To characterize the effect of cross resonance driven at upper sideband frequency of the A mode we freeze the B mode at 0 excitation and look into the two qubit subspace composed of $\left\lbrace 000, 010, 100, 110\right\rbrace $. The desired effective Hamiltonian is a block-diagonal matrix partitioned as $\left\lbrace 000, 010\right\rbrace $, $\left\lbrace 100, 110\right\rbrace $, $\left\lbrace \rm{rest}\right\rbrace $ where the first block has an energy scale of 0 and the second block has an energy scale of $\Delta_{\rm{AT}}$. The following blocks differ by a larger energy scale.
	Now we find a unique unitary matrix $\mathcal{T}$ that block diagonalizes $\hat{\mathcal{H}}_{\rm{sys}}^{\rm{RWA}}$ in the given form.
	\begin{equation}
	\label{BDunitary}
	\hat{\mathcal{H}}_{\rm{eff}}=\mathcal{T}^{\dagger}\hat{\mathcal{H}}_{\rm{sys}}^{\rm{RWA}}\mathcal{T}
	\end{equation}
	This is done by first rotating the given Hamiltonian into its diagonal form by applying the unitary rotation $\mathcal{S}$ defined by the energy eigenvector matrix and then rotate it back using another unitary matrix $\mathcal{F}$ with the additional constraint of block-diagonal form of $\hat{\mathcal{H}}_{\rm{eff}}$. The secular equations for $\hat{\mathcal{H}}_{\rm{sys}}^{\rm{RWA}}$ and $\hat{\mathcal{H}}_{\rm{eff}}$ can be written as:
	\begin{equation}
	\label{secular_eq}
	\begin{split}
	\hat{\mathcal{H}}_{\rm{sys}}^{\rm{RWA}}\mathcal{S}=\mathcal{S}\Lambda\\
	\hat{\mathcal{H}}_{\rm{eff}}\mathcal{F}^{\dagger}=\mathcal{F}^{\dagger}\Lambda
	\end{split}
	\end{equation}
	where $\Lambda$ is the diagonal matrix of eigenvalues $\lambda_i$. Then $\mathcal{T}=\mathcal{S}\mathcal{F}$ and $\hat{\mathcal{H}}_{\rm{eff}}$ will be block diagonalized if $\mathcal{F}$ is a block diagonal matrix of same partitions. Apart from the constraint of the block diagonal form, the matrix $\mathcal{F}$ defined this way and hence the effective Hamiltonian $\hat{\mathcal{H}}_{\rm{eff}}$ is not unique. In order to find the unique effective Hamiltonian we use principle of least action which implies that the transformation matrix $\mathcal{T}$ should change the original matrix as little as possible while making it block diagonal. Mathematically this is satisfied when,
	\begin{equation}
	\label{euclidean_norm}
	||\mathcal{T}-\mathcal{I}||=\rm{minimum}
	\end{equation}
	where $||A||$ is the Euclidean norm and $\mathcal{I}$ is the unit matrix. The resulting solution is given by\cite{least_action},
	\begin{equation}
	\label{least_action}
	\mathcal{T}=\mathcal{S}\mathcal{S}^{\dagger}_{\rm{BD}}\left( \mathcal{S}_{\rm{BD}}\mathcal{S}^{\dagger}_{\rm{BD}}\right)^{-1/2}
	\end{equation}
	where $\mathcal{S}_{\rm{BD}}$ is the block-diagonal part of $\mathcal {S}$ with the same partition as the desired effective Hamiltonian. From the effective Hamiltonian obtained in Eq.\ref{BDunitary} we find the interaction strengths of the resulting components by comparing the matrix elements:
	\begin{equation}
	\label{interaction_matrix}
	\begin{split}
	&\mathcal{H}_{\rm{eff}}=\\
	&\frac{1}{2}(
	\rm{II}\sigma_0\otimes\sigma_0
	+IX\sigma_0\otimes\sigma_1
	+IY\sigma_0\otimes\sigma_2
	+IZ\sigma_0\otimes\sigma_3+\\
	&\rm{ZI}\sigma_30\otimes\sigma_0
	+ZX\sigma_3\otimes\sigma_1
	+ZY\sigma_3\otimes\sigma_2
	+ZZ\sigma_3\otimes\sigma_3)\\ 
	\end{split}
	\end{equation}
	In our numerical calculations we have dropped the exchange coupling between the B mode of dimon and the transmon. This mode is also far detuned in frequency with the transmon compared to the  other mode involved in cross resonance. We observe that under these circumstances, the dynamics is well explained by approximating the A mode energy levels by  a transmon-like system given by $\left\lbrace 00, 10, 20, 30...\right\rbrace $ where we fix the B mode at 0 excitation.
	
	\section{Hamiltonian tomography and Bloch equation}
	Consider a two-level system evolving under the generalized Hamiltonian $\mathcal{H}=\bm{\mathcal{B}}\cdot\bm{\sigma}/2$, where $\bm{\sigma}$ is the Pauli vector and $\bm{\mathcal{B}}=\{\Omega_x, \Omega_y, -\Delta\}$ is the generalized field. $\Omega_{\{x,y\}}$ is the effective drive strength along $\{x,y\}$ axis and $\Delta$ is effective detuning between the drive and the qubit transition frequency inducing Z rotation. We can write the Heisenberg equation of motion as:
	\begin{equation}
	\label{bloch1}
	-i\frac{\partial\rho}{\partial t}=[\rho, \mathcal{H}]=\frac{1}{2}[\bm{r}\cdot\bm{\sigma}, \bm{\mathcal{B}}\cdot\bm{\sigma}]
	\end{equation}
	where $\bm{r}$ is the Bloch vector. Expanding the commutator we get,
	\begin{equation}
	\label{bloch2}
	\dot{\bm{r}}(t)=\mathcal{G}\bm{r}(t)
	\end{equation}
	and matrix $\mathcal{G}$ is given by,
	\begin{equation}
	\label{bloch3}
	\mathcal{G}=\begin{pmatrix}
	0 & \Delta & \Omega_y\\
	-\Delta & 0 & -\Omega_x\\
	-\Omega_y & \Omega_x & 0
	\end{pmatrix}
	\end{equation}
	Integrating EQ \ref{bloch2} we get,
	\begin{equation}
	\label{bloch4}
	\bm{r}(t)=e^{\mathcal{G}t}\bm{r}_0
	\end{equation}

	If we start from the ground state of the two level system, i.e. $\bm{r}_0=\{0,0,1\}$ we get,
	\begin{equation}
	\begin{split}
	\label{bloch_eq}
	\langle x(t)\rangle=&-\frac{\Omega_x\Delta+\Omega_x\Delta\cos(\mathcal{B}t)-\Omega_yR\sin(\mathcal{B}t)}{\mathcal{B}^2}\\
	\langle y(t)\rangle=&-\frac{\Omega_y\Delta+\Omega_y\Delta\cos(\mathcal{B}t)+\Omega_xR\sin(\mathcal{B}t)}{\mathcal{B}^2}\\
	\langle z(t)\rangle=&\frac{\Delta^2+(\Omega_x^2+\Omega_y^2)\cos(\mathcal{B}t)}{\mathcal{B}^2}
	\end{split}
	\end{equation}
	where, $\mathcal{B}=\sqrt{\Omega_x^2+\Omega_y^2+\Delta^2}$. 
	
\begin{figure}[t]
		\centering
		\includegraphics[width=0.5\textwidth]{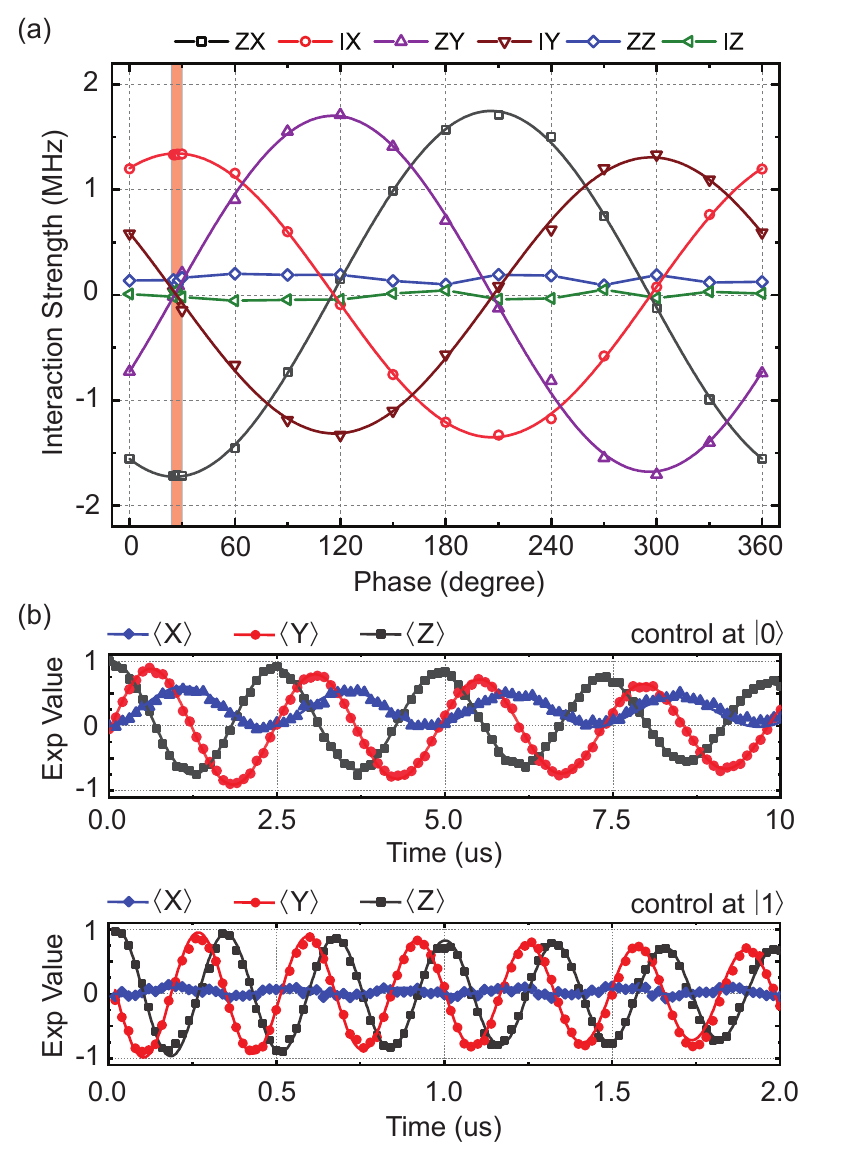} 
		\vspace{0pt}
		\caption{(a) Effective strengths of different interaction terms in the Hamiltonian as a function of drive phase. The phase of the CR drive is adjusted such that the ZY interaction is minimized (area highlighted in light orange). We observe that the same drive phase also cancels the IY term, indicating negligible classical cross-talk in this architecture. (b) Evolution of the target qubit (A mode of dimon) with the control qubit (transmon) at ground and excited state at the optimum drive phase. The evolution is fit to the Bloch equation with an exponential decay factor. Comparing the fit parameters from the two sets of oscillations we estimate the interaction strengths in the Hamiltonian.}
		\label{fig:ht}
	\end{figure}

	Hamiltonian tomography consists of two sets of experiments where we keep the control qubit in $|0\rangle$ and $|1\rangle$ states respectively and apply the cross resonance drive on the control qubit for a variable time ranging from $0$ to $\tau_{\rm{max}}$. Then we perform a tomographic measurement on the target qubit and plot the evolution of the expectation values $\langle x(t)\rangle_{\rm{T}}$, $\langle y(t)\rangle_{\rm{T}}$ and $\langle z(t)\rangle_{\rm{T}}$ for the target qubit. These oscillations are fit to Eq. \ref{bloch_eq} with an additional decay factor due to finite coherence. The interaction strengths $\Omega_x^{\{0, 1\}}, \Omega_y^{\{0, 1\}}$ and $\Delta^{\{0, 1\}}$ are extracted from the fitting corresponding to the control qubit states $|0\rangle$ and $|1\rangle$. From sum and difference of these parameters we find the six possible interaction terms in the effective block diagonal CR Hamiltonian viz. ZX, ZY, ZZ, IX, IY and IZ, e.g. $\rm{ZX}=$$(\Omega_x^0-\Omega_x^1)/2$ and $\rm{IX}=$$(\Omega_x^0+\Omega_x^1)/2$ etc. A typical set of evolution in Hamiltonian tomography experiment is shown in Fig.~\ref{fig:ht}(b).
	
	To tune the phase of CR drive with respect to the target qubit axes we perform several Hamiltonian tomography experiments with varying phase of the applied drive tone. The dependence is shown in Fig.~\ref{fig:ht}(a). We observe that the ZY and IY term simultaneously goes to zero when the phase is properly aligned, implying that there is almost no classical cross talk in this geometry.
	
\begin{figure}[t]
		\centering
		\includegraphics[width=0.5\textwidth]{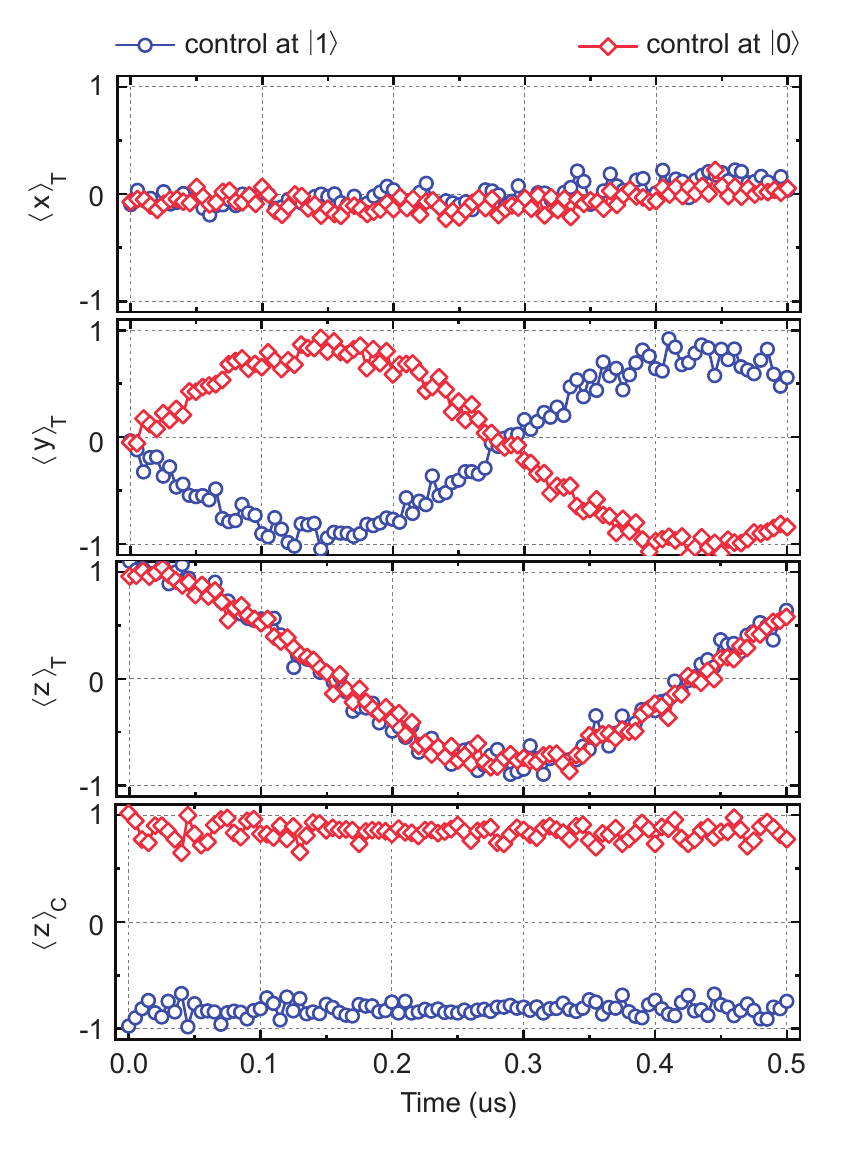} 
		\vspace{-10pt}
		\caption{Qubit state evolution under echoed cross resonance drive. The sequence consists of two  CR/2 pulses of variable lengths with a fixed 10ns Gaussian rise and fall time. The two CR/2 pulses are of opposite phases and are separated by a $\pi$-pulse on control qubit. The expectation values of X, Y and Z operators of the target qubit (Dimon A mode) are plotted in figure (a), (b) and (c) respectively. In figure (d) the Z projection of the control qubit (transmon) is plotted. When the phase is properly calibrated the oscillations in X projection should be minimized. We observe a residual oscillation in X projection due to ZZ term present in the Hamiltonian. There are also small oscillations in the control qubit that limits the maximum drive power of the cross resonance pulse. These oscillations are due to strong off resonant drive on the control qubit leading to leakage. The shaded region marks the gate time for calibrated $\rm{ZX}_{\pi/2}$ operation.}
		\label{fig:ecr}
	\end{figure}

	Finally an echoed cross resonance is performed with the calibrated drive phase. The oscillations are plotted in Fig~\ref{fig:ecr}. We observe a weak oscillation in the measured value of $\langle x\rangle$ due to residual ZZ error. We estimate the gate times from the oscillation in $\langle y\rangle$ and $\langle z\rangle$. We have also plotted the evolution of the control qubit under the CR drive. We observe fast small oscillation in the control qubit due to the presence of the off resonant drive. These oscillations indicate leakage out of subspace and limit the speed of cross resonance gate beyond certain amplitude.

	\section{Quantum Process Tomography}
	A two qubit quantum gate $\varepsilon$ acting on an input density matrix $\rho_{in}$ can be represented as  \cite{kraus}\cite{QPT1}
	\begin{equation} \label{eq:1}
	\varepsilon(\rho_{in})=\sum\limits_{i=1}^{d^{2}}E_{i}\rho_{in} E_{i}^\dagger
	\end{equation}
	where $\varepsilon(\rho_{in})$ is the output density matrix, $d=2^{n}$ is the dimension of Hilbert space ($n$ =no. of qubits) which in this case is 4. [$ E_{i}$] are known as Kraus operators satisfying the completeness relation $\sum_{i}E_{i}^\dagger E_{i}=\mathbb{I}$.
	
	Quantum process tomography (QPT) involves determining these operators [$E_{i}$] \cite{QPT_define} \cite{yamamoto}. Each of these Kraus operators can again be represented by a fixed set of basis operators [$A_{i}$] such that $E_{i}=\sum_{m}e_{im}A_{m}$. Substituting this back in equation \ref{eq:1} we get
	\begin{equation}\label{eq:2}
	\rho_{\rm{out}}=\varepsilon(\rho_{\rm{in}})=\sum\limits_{m,n}^{16} \chi_{mn}A_{m}\rho_{\rm{in}} A_{n}^\dagger 
	\end{equation}
	where, $\chi_{mn}=\sum_{k}e_{mk}e_{kn}^{*}$ are elements of a positive Hermitian matrix called process matrix $\chi$. This process matrix completely characterizes the quantum gate, since it describes how much $A_{m}\rho_{\rm{in}} A_{n}^\dagger$ contributes to $\rho_{\rm{out}}$.
	
	\begin{figure}[t]
		\centering
		\includegraphics[width=0.5\textwidth]{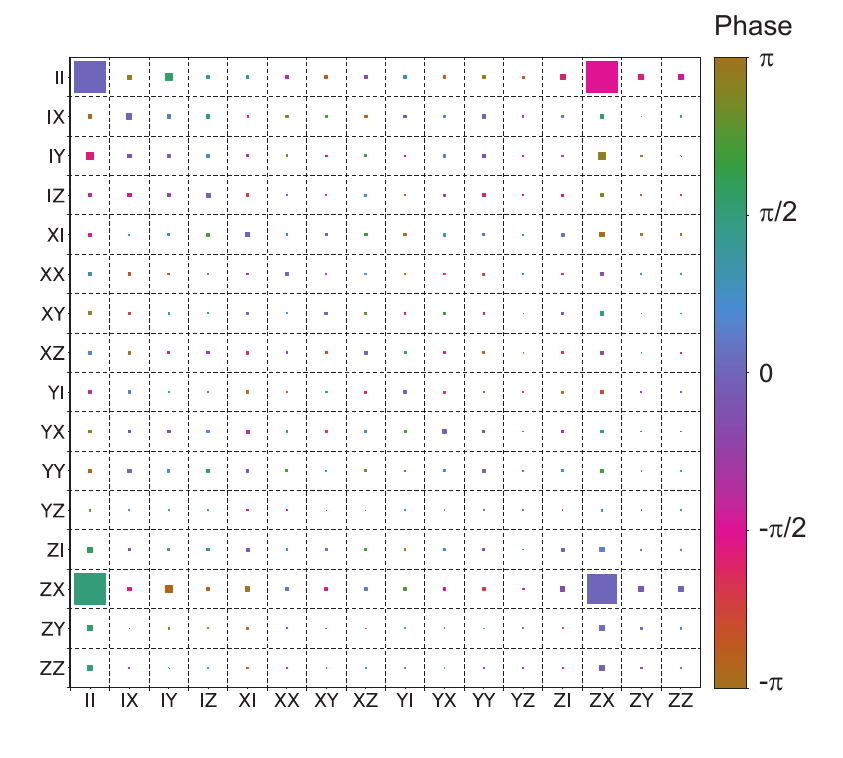} 
		\vspace{-10pt}
		\caption{Process matrix of the $\rm{ZX}_{\pi/2}$ gate estimated by quantum process tomography on the two-qubit system. The system is prepared in 16 input states and then the cross resonance gate is applied with echo sequence. Finally a quantum state tomography is performed on the result state. Qubit states are initialized with heralding to ground state. The final states are corrected for measurement error. We report a process fidelity of 0.9282 and a corresponding gate fidelity of 0.9426 in our implementation of $\rm{ZX}_{\pi/2}$ operation.}
		\label{fig:qpt}
	\end{figure}
	
	To experimentally reconstruct the process matrix $\chi$ by QPT , we first prepare $d^{2}=16$ linearly independent input density matrices [$\rho_j$] chosen from states ($|0\rangle$,$|1\rangle$,$|0\rangle+|1\rangle$ and $|0\rangle-i|1\rangle$) for each qubit. Operating our two-qubit gate on these states, we reconstruct the final density matrices $\varepsilon(\rho_j)$ by performing quantum state tomography (QST). These output density matrices are written as 
	\begin{equation}\label{eq:3}
	\varepsilon(\rho_{j})=\sum\limits_{k=1}^{16} \lambda_{jk}\rho_{k}
	\end{equation}
	From the measurement results of {$\varepsilon(\rho_{j})$}, the coefficients $\lambda_{jk}$ are determined. 
	Now equating equations \ref{eq:2} and \ref{eq:3} one can write 
	\begin{equation}
	\varepsilon(\rho_{j})=\sum\limits_{k} \lambda_{jk}\rho_{k}=\sum\limits_{m,n,k} \chi_{mn} \beta_{jk}^{mn}\rho_{k}
	\end{equation}
	where we define the coefficients $\beta_{mn}^{jk}$ such that 
	\begin{equation}
	A_{m}\rho_{j} A_{n}^\dagger=\sum\limits_{k} \beta_{jk}^{mn}\rho_{k}
	\end{equation}
	For two-qubits, we choose $A_{m}$ = $B_{i}$ $\otimes$ $B_{j}$ from a fixed set of single-qubit basis operators $B_{i} \in \{I,X,Y,Z\}$ \cite{naik_random}. The process matrix $\chi_{mn}$ is then related to the lambda matrix $\lambda_{jk}$ as 
	\begin{equation}
	\lambda_{jk}=\sum_{mn} \chi_{mn} \beta_{jk}^{mn}
	\end{equation}
	The above equation can be directly inverted to obtain the $16\times16$ process matrix $\chi_{mn}$. 
	
	In presence of experimental noise, this process matrix $\chi^{exp}$ obtained as above is not necessarily physical, i.e, it can be non-Hermitian and have negative eigenvalues. Therefore, we find the physical matrix $\chi^{p}$ which is closest to the experimental process matrix $\chi^{\rm{exp}}$ \cite{CNOT_QPT} \cite{bialczak}by minimizing the following function \cite{iop}
	\begin{equation}
	f(t)=||\chi^{\rm{exp}}-\chi^{p}(t)||_{2}+\lambda(\sum\limits_{m,n}^{16} 	\chi^{p}_{mn}(t)A_{m}\rho_{in} A_{n}^\dagger-\mathbb{I})
	\end{equation}
	Here, we parameterize the physical process matrix as
	\begin{equation}
	\chi^{p}(t)=\frac{T(t)^\dagger T(t)}{{\rm{tr}}[T(t)^\dagger T(t)]}
	\end{equation}
	where $T(t)$ is a $16 \times 16$ complex lower triangular matrix with 64 real parameters $t_{j}$. 
	\begin{equation}
	T(t)=\begin{bmatrix} 
	t_{1} & 0 & 0 & 0 & \dots & 0\\
	t_{17}+ i t_{18} & t_{2} & 0 & 0 & \dots & 0\\
	t_{33}+ i t_{34} & t_{19}+ i t_{20} & t_{3} & 0 & \dots & 0\\
	\vdots & \vdots & \ddots & \vdots & \vdots & \vdots\\
	t_{63}+ i t_{64} & \dots & \dots & \dots & \dots & t_{16}
	\end{bmatrix}
	\end{equation}
	and $\lambda$ is a Lagrange multiplier used for ensuring trace preservation for $\chi^{p}(t)$. The algorithm used for minimizing the above function $f(t)$ was Nelder-Mead algorithm similar to the process used in \cite{iop}.
	
	We then compare our physical process matrix $\chi^{p}$ with the ideal process matrix $\chi^{id}$ for our two-qubit gate by defining the process fidelity as 
	\begin{equation}
	\mathcal{F}(\chi^{id},\chi^{p})=tr(\chi^{p}\chi^{id\dagger})
	\end{equation}

	\section{Decoherence in copper cavity}
	In the first experiment we tune the transmon frequency using an external dc flux through the squid loop. In 3D cQED this is usually done by placing the split junction transmon in a copper cavity attached to a superconducting solenoid magnet. We observe that the tunable transmon and dimon suffered from poor coherence times in the copper cavity. Possible explanation for the bad coherence is the participation of lossy copper in the qubit capacitance. In the bridge geometry the qubit chip is placed across two cavities through a narrow hole surrounded by copper where the electric fields of the qubit dipoles are highly concentrated. Coherence of the same qubits improve when we remeasured it in an aluminum cavity of identical geometry. We also measured improvement in the coherence times when we put thin strip of indium on top of the copper layer near the qubits, creating a superconducting environment around it. The coherence can improved further by using techniques to apply DC magnetic field inside superconducting cavities\cite{magnetic_hose, reed}.

	\section{Limit set on gate fidelity due to finite coherence}
	The cross resonance gate calibrated in the main text suffers from both small unitary error due to ZZ coupling and incoherent processes like relaxation and dephasing. We put a theoretical bound on the gate fidelity due to finite coherence of the qubits.
	For single qubit, the gate error due to decoherence, $\epsilon_1=(1-F_g)$ is given by\cite{coherence_limit},
	\begin{equation}
	\label{single_qubit_err}
	\epsilon_1=\frac{1}{2}\left[1-\frac{2}{3}\rm{exp}\left(-\frac{\tau_g}{\rm{T2}}\right)+\frac{1}{3}\rm{exp}\left(-\frac{\tau_g}{\rm{T1}}\right)\right]
	\end{equation}
	where, $\rm{T1}$ is the relaxation time and $\rm{T2}$ is the spin echo time.
	For two qubit gates, the gate error due to decoherence is estimated by the following formula:
	\begin{equation}
	\label{two_qubit_err}
	\epsilon_2=\frac{3}{4}\left(1-\zeta_{\rm{T1}}-\zeta_{\rm{T2}}\right)
	\end{equation}
	where,
	\begin{equation}
	\label{t1_factor}
	\zeta_{\rm{T1}}=\frac{1}{15}\left(\rm{exp}\left(-\frac{\tau_g}{\rm{T1_{Q1}}}\right)+\rm{exp}\left(-\frac{\tau_g}{\rm{T1_{Q2}}}\right)\right)
	\end{equation}
	and 
	\begin{equation}
	\label{t2_factor}
	\begin{split}
	\zeta_{\rm{T2}}=&\frac{2}{15}\left[\rm{exp}\left(-\frac{\tau_g}{\rm{T2_{Q1}}}\right)+\rm{exp}\left(-\frac{\tau_g}{\rm{T2_{Q2}}}\right)\right]\\
	+&\frac{2}{15}\rm{exp}\left(-\tau_g\left(\frac{1}{\rm{T2_{Q2}}}+\frac{1}{\rm{T1_{Q1}}}\right)\right)\\
	+&\frac{2}{15}\rm{exp}\left(-\tau_g\left(\frac{1}{\rm{T2_{Q1}}}+\frac{1}{\rm{T1_{Q2}}}\right)\right)
	\end{split}
	\end{equation}
	$\rm{Q1}$ and $\rm{Q2}$ in the subscript represent the two qubits involved in the multi-qubit gate.

\end{document}